Fine scale depth regulation of invertebrate larvae around coastal fronts

Nicolas Weidberg (1,2,*), Wayne Goschen (3, 4), Jennifer M. Jackson (2,5), Paula Pattrick (2,3), Christopher D. McQuaid (1), Francesca Porri (1,2)

1 Department of Zoology and Entomology, Rhodes University, Grahamstown 6140, South Africa.

2 South African Institute for Aquatic Biodiversity, Somerset Street, Grahamstown 6139, South Africa.

3 South African Environmental Observation Network, Elwandle Coastal Node, Nelson Mandela University Ocean Sciences Campus, Port Elizabeth 77000, South Africa.

4 Institute for Coastal and Marine Research, Nelson Mandela University, Port Elizabeth 77000, South Africa

5 Hakai Institute, Victoria BC, Canada.

* present address: Department of Arctic Marine Biology, University of Tromso, Tromso 9019, Norway.

Emails: Nicolas Weidberg, j_weidberg@hotmail.com, nlo009@uit.no

Wayne Goschen, wayne@saeon.ac.za

Jennifer Jackson, jennifer.jackson@hakai.org

Paula Pattrick, p.pattrick@saiab.ac.za

Christopher D. McQuaid, c.mcquaid@ru.ac.za

Francesca Porri, f.porri@ru.ac.za

Abstract

Vertical migrations of zooplankters have been widely described, but their active movements through shallow, highly dynamic water columns within the inner shelf may be more complex and difficult to characterise. In this study, invertebrate larvae, currents and hydrographic variables were sampled at different depths during and after the presence of fronts on three different cruises off the southern coast of South Africa. Internal wave dynamics were observed in the hydrographic dataset but also through satellite imagery, although strong surface convergent currents were absent and thermal stratification was weak. During the first two cruises, fronts were more conspicuous and they preceded strong onshore currents at depth which developed with the rising tide. Vertical distributions of larvae changed accordingly, with higher abundances at these deep layers once the front disappeared. The third cruise was carried out during slack tides, the front was not conspicuous, deep strong onshore currents did not occur afterwards and larval distributions did not change consistently through time. Overall, the vertical distributions of many larval taxa matched the vertical profiles of shorewards currents and multivariate analyses revealed that these flows structured the larval community, which was not influenced by temperature nor chlorophyll. Thus, the ability to regulate active vertical positioning may enhance shorewards advection and determine nearshore larval distributions.

Introduction

Diel vertical migrations of zooplankters across the world's oceans have been studied since the development of sonar in the 20th century revealed the magnitude of this phenomenom (Barham 1963, Heywood 1996, Fornshell and Tesei 2013). The main selective pressures thought to cause these 24h vertical displacements are visual predation and metabolic rates as both would be too high in clear, relatively warm surface waters during the day (Enright 1977, Stich and Lampert 1981). Although widespread throuhghout the planktonic community, diel vertical migration is known to be highly variable and responsive to organismal physiological condition (Hays et al. 2001) and prevailing environmental forcing (Weidberg et al. 2015), thus it can be combined with other active behaviours across the water column. When studied in shallower waters over the continental shelves, however, other more complex patterns of vertical movements emerged. For coastal meroplankters, most patterns seem related to offshore advection, which is an important selective pressure on pelagic organisms specific to coastal regions (Jackson and Strathmann 1981). For coastal plankton, and especially for the larvae of coastal benthic invertebrates, offshore transport may be a source of mortality and could result in recruitment failure, where not enough larvae settle to replace the adult population. Thus, by swimming against the weak vertical flows imposed by upwelling-downwelling transitions and other hydrographical processes, coastal plankton can remain at depths where currents are directed onshore (Genin et al 2005, Queiroga and Blanton 2004, Shanks and Shearman 2009).

The ability to regulate their depth allow coastal plankton and larvae to respond to nearshore hydrographical structures such as fronts. Fronts can be defined as oceanographic features that separate two different water types, and are typically characterised as having a vertical current structure. Many different invertebrate larvae have been observed to accumulate at fronts by first being horizontally transported to them and then by swimming against the downward vertical flow which characterises these convergent structures (Franks 1992, Pineda 1999, Shanks et al. 2000, Weidberg et al. 2014). Fronts can be formed by multiple hydrographical processes operating at different spatio-temporal scales including Ekman forcing, estuarine circulation, tides and topographically driven currents (LeFevre 1986, Shanks et al. 2003). Processes that cause accumulation at these fronts, like internal waves and upwelling, may become crucial for benthic larval aggregation. In particular, convergent features produced by internal wave fronts and their associated bores have been shown to promote the aggregation and shorewards transport of invertebrate larvae (Pineda 1999, Helfrich and Pineda 2003, Weidberg et al. 2014). These internal motions are usally caused by tidal forcing over abrupt topographies at the shelf break which causes instabilities that travel along the thermocline (Pond and Pickard 1983, Holloway 1987). The potential onshore displacement of larval accumulations driven by internal waves may have profound consequences for the final recruitment of individuals into the adult populations as suggested by the positive association of frontal occurrence and larval settlement (Lagos et al. 2008, Woodson et al. 2012). It is, however, difficult to establish the importance of fronts on larval dispersal as it may depend on their frequency, persistence, strength and seasonal interaction with larval biological features (Largier 1993). In addition, in the case of internal tidal waves there might also be shoreward transport at depth, when the typical foam lines and slicks are not visible at the surface (Pineda 1991, Lievana MacTavish et al. 2016).

Along the south coast of South Africa (Figure 1) there are a variety of processes which are known to generate fronts at many different scales, times and locations. A major mesoscale frontal feature separates the shelf Agulhas Bank waters from the warm core of the Agulhas Current (Lutjeharms 2006). In addition, Ekman forcing is not the only source of upwelled waters in the region: fast flow velocites typical of the Agulhas Current and its meanders onto the shelf cause shelf edge upwelling, semipermanent upwelling cells and their associated fronts (Schuman et al. 1982, Goschen and Schuman 1990, Lutjeharms et al. 2000, Goschen et al. 2015). Internal waves have also been observed across the Agulhas Bank (Largier and Swart 1987, Jackson et al. 2014). All these structures can potentially affect the distributions of a wide invertebrate larval community with highly adaptive active responses. Fronts have been hypothesised to influence the distribution of mussel larvae in the region (McQuaid and Phillips 2000) and high abundances were associated with internal wave fronts over the Agulhas Bank (Porri et al. 2014, Weidberg et al. 2015).

In this study, we sampled larval distributions around frontal structures within the inner shelf waters off the south coast of South Africa. Our sampling methodology allowed the physical characterisation of the events in an attempt to identify the physical drivers behind their development. In addition, as we sampled during and after frontal occurrence, we were able to quantify for the first time the persistence of frontal larval distributions, a key feature that can allow us to evaluate the long term effects of coastal fronts on larval transport, settlement and population dynamics.

## Materials and Methods

### Field samplings

On three different days during the austral spring of 2014 (25 September, 6 October and 29 October, hereafter Events 1, 2 and 3, respectively), nearshore waters off Sardinia Bay (34.03-34.07°S, 25.47-25.51°E) on the western exposed side of Cape Recife (Fig. 1) were surveyed in search of foam lines and/or oily slicks parallel to the coast on board the 13m long R/V uKwabelana. In addition, LANDSAT imagery was also retrieved for the same days at the area surveyed to observe the same surface structures (Supplementary Information 1). Once the front was located by observing foam lines or oily slicks, a total of 6 stations were set. These were arranged as two cross-frontal transects (A and B), each with 1 station at the front, 1 between the front and the shore (250-700m from the front) and 1 on the seaward (at 200-700m from the front) side of the front. These stations were defined as frontal, onshore and offshore stations, respectively. At each station, an SBE19 plus CTD with a Wetlabs ECO- AFL fluorometer was lowered from the surface to the bottom and the temperature profiles obtained were inspected in search of a rapid drop in temperature with depth (approximately 1°C/5m) pointing to the presence of a thermocline. A 2.2 KC Denmark 23.580 plankton pump with a net of 60 micron mesh size was deployed three times; once at 1m depth, once at the thermocline (or at mid depths in the absence of a thermocline), and once 2-3m above the bottom to collect the biological larval samples. The pump filtered at each depth for about 7 minutes which yielded a mean volume of 1.3$m^3$ (between 0.9 and 2.3 $m^3$). At the same time, an Acoustic Doppler Current Profiler (ADCP) Teledyne RD Instruments Workhorse Monitor of 300kHz frequency, beam angle 20 degrees and a 4 beam system was deployed facing downward, at 0.5m depth, on the port side of the boat for a mínimum of 30 minutes. The instrument registered currents in three axes (E-W, N-S and vertical) every 0.5m by emitting one ping every 1.5 seconds and using 1 ping to calculate one individual measurement or ensemble. It also recorded bottom tracking to account for boat drift. The 6 stations sampled took about 3-4.5 hours (Time 1, front present). Once the front dissipated the same sampling design was repeated at or close to the initial stations (3.5-4.5 hours, Time 2, front dissipated).

Physical variables

Cross shore profiles for temperature and chlorophyll-a were obtained from the CTD and fluorometer sensors at each transect for each event during and after the presence of the front. CTD and fluorometer data were processed with Seabird's Seasoft V2 software, using the standard steps recommended for the SBE 19 Plus. ADCP profiles were collected at the same stations and these data were processed. The processing steps used included substracting bottom tracking velocities, discarding data with less than 100% good criteria, discarding data where the velocity error was greater than 40 $cm*s^{-1}$ and discarding estimates of mean currents calculated with less than 100 counts per 0.5m depth cell. Currents were then averaged over the total time of deployment at a given station (between 30 and 45 minutes).

Hourly measurements of wind direction and velocity during the cruises were obtained from the South African Weather Service (SAWS) at Port Elizabeth International Airport, 12km to the northeast. Speeds were corrected for height using the wind profile power law (Hsu et al. 1994) to get estimates at 9m following the same procedure applied for the region in previous studies (Weidberg et al. 2015). From these data, an upwelling index as zonal Ekman transport was calculated (Bakun 1973) with positive and negative values to indicate offshore and onshore displacements of surface waters, respectively. In addition, tidal motions were retrieved at Port Elizabeth with the programme WXTide32 with a time resolution of 10 minutes. Tidal ranges in the region span approximately 0.4 to 2m.

Plankton samples

The cod end of the plankton pump was rinsed with filtered sea water and its contents was poured into 250ml plastic jars into which 95% ethanol was added. These samples were inspected in the laboratory with a microscope (Zeiss Stemi DV4) to identify the larvae of benthic invertebrates. We identified all larvae to the lowest taxonomic category possible, family and stages in the case of barnacles. Organisms were counted through the whole sample and their numbers were divided by the total volume filtered to standardise abundances as individuals/$m^3$.

Frontal circulation

Given the orientation of the coast, the North-South axis was considered a good proxy of shorewards-seawards directions. Thus, from the velocity estimates obtained with the ADCP, the cross shore-meridional currents at those depths at which plankton abundances were sampled were used to infer accumulation speeds. Subsequently, accumulation speeds (velocities at which particles may accumulate at either side of the front, see Pineda 1999 and Weidberg et al. 2014 for details) were calculated as:

$\Delta Von = Vf-Von$

$\Delta Voff = Voff-Vf$

where Von, Voff and Vf are cross shore speeds at onshore, offshore and frontal waters, with positive values indicative of northwards, onshore direction. These were vertically averaged for each 0.5 m binned depth, horizontally averaged for each onshore, frontal, and offshore pair of stations. These calculations were done during and after frontal development (Times 1 and 2, respectively) thus enabling the estimation of a net difference between times for both the onshore and offshore sides of the front as:

$\Delta Von\ (T2-T1) = \Delta VonT2 - \Delta VonT1$

$\Delta Voff\ (T2-T1) = \Delta VoffT2 - \Delta VoffT1$

Statistics

In order to infer the accumulation patterns of larvae around fronts, the abundances of the most abundant taxa (those with a minimum of 3 ind*m$^{-3}$ at one station at a given depth) were used as dependent variables in factorial analyses of variance (ANOVA), with position (onshore, front, or offshore) and depth (surface, middle, or bottom) as fixed factors. As variances of data for the interaction between position and depth were often heterogeneous according (Levene´s test $P < 0.05$), abundances were log transformed. To ensure that samples were independent within each ANOVA, for each taxon and event, separate analyses were carried out during frontal occurrence (Time 1) and after frontal dissipation (Time 2).

A proxy of larval vertical positioning in the water column was calculated as the mean larval depth (MLD; Tapia et al. 2010), which is a mean of depths sampled at a given station weighted by the respective larval abundances:

$MLD = \sum(N_i * D_i) / \sum N_i$

where $N_i$ is larval abundances at depth $D_i$ (surface, middle or bottom) and $\sum N_i$ is the total number of larvae at the station. These calculations were done at stations where larvae reached at least 3 ind/m³ at one depth. Similarly, we also obtained a mean depth weighted by cross shore currents V (MVD) as

$$MVD=\sum(V_i*D_i)/\sum V_i$$

where $V_i$ is cross shore flows at depth $D_i$ (from the surface to the bottom every 0.5m) and $\sum V_i$ is the summation of flows through the water column. For these calculations, flows were re-scaled by substracting the maximum negative value (i.e. the fastest seawards current) from all values at a given station so that all resultant values were positive. Then, to infer how the vertical dynamic structure of the cross shore flow affected vertical larval distributions, MLDs were regressed againts MVDs by means of ordinary least square (OLS) regression for the most abundant taxa. The variance explained by these regressions ($R^2$) was used as a proxy of the degree of coupling between larval and cross shore flows distributions for a specific taxon.

Multivariate principal component analyses (PCA) were performed from averages of the most important physical variables, after normalization, for every larval taxon: temperature (T), chlorophyll-a (CHLA), depth (D) and velocities for the three axes: North-SouthU; East-West (V) and vertical (Z). This procedure allowed insight into which physical variables most influenced the structure of the larval community. Three different PCAs were carried out with means of the physical variables calculated for all samplings (Time1 plus Time 2); for sampling during Time 1 only; and those done during Time 2 only. Correlations between physical variables were examined for each PCA and if a pair of variables presented a Pearson´s R value greater than 0.8, one of them was removed from the analysis to avoid excesive multicollinearity (Clarke and Gorley 2006). The Pc1 of these analyses was considered as a descriptor of larval assemblages and its taxon specific values were correlated with the $R^2$ of the relationships between MLDs and MVDs. In addition, larval swimming velocities obtained from the literature (Chia et al. 1974, Weidberg et al. 2014) were represented for all multivariate analyses to establish any potential relationship with larval active locomotion.

Results

Field sampling

The frontal structures observed off Sardinia Bay in 2014 on 25 September, 6 October and 29 October (hereafter Events 1, 2 and 3, see Figure 1) shared some visual characteristics. On the first two dates, several surface slicks parallel to the shore were observed and the most conspicuous and closest to shore was sampled. This pattern was especially evident during Event 2, with 5 consecutive narrow foam structures approximately 100 m apart observed at the beginning of the sampling period. These foam structures were confirmed by LANDSAT images (Supplementary Information 1). In addition, fish activity was observed at the surface within the slicks and Sandwich terns (*Thalasseus sandvicensis*) were flying around the front (Pers obs). During Event 3, a single 100m wide slick was observed which did not move shorewards and did not present any foam.

Physical variables

Wind forcing was relatively weak during our cruises, with winds never exceeding velocities of 7m*s$^{-1}$ or absolute Ekman transport rates of 1000 m$^3$ km$^{-1}$ s$^{-1}$ (Fig. 2). This was especially evident for Events 1 and 2, while during Time 2 (front dissipated) in Event 3 stronger downwelling, westerly winds blew (Fig. 2). Spring tides occurred during Events 1 and 2, sampled 1 day after new moon and 2 days before full moon, respectively, with ranges greater than 1.5m, while Event 3 occurred 2 days before the first quarter moon corresponding to a neap tide with a range of around 1m. Fronts consistently occurred as tides were close to their lowest levels

(during flood tides for Events 1 and 2 and during ebb tide for Event 3) while they disappeared as the tide was rising (Fig. 2).

Cross shore contour profiles of chla, temperature and currents revealed differences in the structure of the water column between events and times but also some similaritites. During Event 1, the warmest waters were recorded close to the surface (17.4°C) and relatively cold waters (~14°C) were offshore at the bottom, especially during Time 1 (front present). The limited vertical differences in temperatures were not enough to form a marked thermocline. Chlorophyll-a values were quite low, close to normal winter values for the region, never exceeding 2 mg*m$^{-3}$. Offshore, currents around 20m depth were flowing westwards and downwards, especially at transect A during Time 1 at very high velocities up to 90 cm*s$^{-1}$, while they shifted northwards during Time 2. Onshore, flows were much weaker and did not present any clear directional pattern (Fig. 3).

Event 2 was characterised by lower temperatures than during Event 1, with a narrow offshore surface layer (10m width) of roughly 14°C that was closer to the shore at Time 2. Onshore waters around 12°C were mixed. Chlorophyll-a values were even lower than on Event 1 and very close to 0, with slightly higher values at mid depths. Again, strong flows at 20-25m depth developed offshore, but this time they were directed eastwards, northwards and upwards at speeds of 20-35cm*s$^{-1}$ during Time 1 (front present). Closer inshore, strong bottom westwards currents developed during Time 1. During Time 2 (front dissipated), currents were flowing mainly shorewards at higher speeds (50cm*s$^{-1}$),

During Event 3, temperatures were in general warmer than during Event 2, with a 10m depth surface layer of 16°C and bottom waters of 12°C. At transect B, slightly colder (Time 1) and warmer (Time 2) waters occupied the whole water column at the front. Chlorophyll-a levels were much higher than during previous events, attaining values of 6mg*m$^{-3}$ but they did not present a clear spatial pattern. Currents were much weaker than previously recorded and the strong offshore deep currents were not present. Instead, relatively strong westwards currents (20cm*s$^{-1}$) were flowing offshore at the surface during Time 1, but this pattern was not clear by Time 2. Bottom waters at the front were flowing seawards at similar speeds to Time 1. No clear structures were obseverd for vertical flows during this event (Fig. 5).

Frontal circulation

Positive relative speeds from both sides of the front, indicative of potential transport towards the front and hence possible particle accumulation, were not particularly faster at the surface than the bottom during frontal occurrence at Time 1 compared to Time 2 (Table 1). In fact, negative relative speeds occurred at some depths during the three events, with no clear temporal or spatial structure. The only clear pattern occurred at the bottom layer for Events 1 and 2, when, at Time 2, positive and stronger shorewards relative speeds developed, leading to the hightest temporal differences in accumulation speeds from both sides of the front (between 15 to 48 cm*s$^{-1}$, Table 1). Thus, for Events 1 and 2, the weakening of the front might be related to the appearence of stronger bottom landwards flows.

Plankton samples and statistics

A wide diversity of zooplankters was found over the three events. Barnacle larvae were separated into two broad taxonomic categories as either balanids (mainly comprising the species *Tetraclita serrata, Balanus glandula* and some megabalanids) or chthamalids (*Octomeris angulosa* and *Chthamalus dentatus*). These larvae were also separated into different developmental stages, early nauplii (nauplii I-III), late nauplii (IV-VI) and cyprids. Within balanid cyprids, two different morphs were found, but could not be taxonomically identified, so they were named as balanid cyprids A and B. *Octomeris angulosa* could only be identified and separated from *Chthamalus dentatus* as late nauplii. Even broader taxonomic categoríes were chosen for the rest of the groups.

Decapod stages were also identified as zoeae and megalopae. Among them, there were pinnotherid and porcelanid zoeae, probably belonging to either *Porcellana* or *Pisidia*. Spionid annelids, possibly of the genus *Polydora*, could be identified and separated from other segmented annelid larvae. Bryozoan cyphonauts identfied could have belonged to the *Membranipora*. Among the echinoderms, pluteus larvae of brittle stars and sea urchins could be separated.

Temporal variability in larval composition was obvious, not only between events, but also between times within events, with some taxa appearing or dissappearing from coastal waters within a few hours (Table 2). Specific analyses of variance for each taxon occurring at each event and time revealed that depth and the interaction depth*position significantly affected larval abundances more than position of the front on its own, especially for Events 1 and 2 and Time 1 (Table 2). Most of the taxa tended to be present at higher abundances at mid-bottom layers and offshore (Table 2, Fig 6 A and C), with the clear exception of bryozoan cyphonauts which were found in greater abundances onshore and at the surface (Table 2, Fig. 6B). Interestingly, during Events 1 and 2, many taxa were located at deeper depths (middle, Event 1, and bottom, Event 2; see Table 2) during Time 2 compared to their distributions at Time 1. This temporal shift was evident for balanid cyprids, late and early nauplii, chthamalid early nauplii and gastropods during Event 1; and for balanid cyprid A and early nauplii, zoeae of anomurans, pinnotherids and other brachyurans, annelids and gastropods during Event 2 (Table 2). During Event 3, while the front was present (Time 1), many taxa were abundant offshore (balanid cyprid A and late nauplii, chthamalid early nauplii, *Octomeris angulosa* late nauplii and anomuran zoeae) and/or were located at the bottom (all decapod taxa), but all these patterns disappeared at Time 2 (Table 2).

Correlations between mean larval depths and mean cross shore flow depths differed across taxa from non-significant with low $R^2$ values in the case of bryozoans to very significant fits explaining up to 70% of variability in mean larval depth (Fig. 7, Fig. 10). Overall, 10 out of 13 different larval taxa showed significant correlations, while there was not a clear association in chthamalid early nauplii, pinnotherid zoeae and bryozoan cyphonauts. Larval and cross shore currents vertical profiles were found to match at some stations for those taxa presenting high $R^2$, especially at Time 2 when strong shorewards currents were present (Fig. 7).

PCA multivariate analyses for all data, Time 1 and Time 2 datasets presented similar patterns, with some variations. The larval community was structured along Pc1 for all analyses, which explained between 51.1, 55.3 and 58.8% of total variabilty of all data, Time 1 and Time 2 analyses, respectively (Fig. 8). The variable that contributed the most to PC1 was always mean cross shore speeds V, while the role of temperature, chl-a, depth and the other vectorial components of current velocities was not as important. V contributed more to Pc1 when data after frontal development were considered (-0.522 compared to -0.513 for all data and -0.475 for Time 1, Fig.8). Pc2 explained much lower percentages of variability (not more than 21.5%) and was associated mainly with the vertical component of flow Z and chl-a (Fig. 8). The range of larval velocities spanned two orders of magnitude, from slow echinoderm pluteus larvae that typically swim 0.01 cm*s-1 to very fast barnacle cyprids that reach several cms*s-1 (Chia et al. 1984). PCA did not however show any effect of larval velocity on larval community structure.

For Time 2, depth had to be discarded as an input variable as it was strongly and positively correlated with V and presented a correlation coefficient greater than the recommended 0.8 (Clarke and Gorley 2006). Thus, taxa located deeper in the water column after frontal development experienced stronger shorewards flows (Fig. 9C). This pattern was blurred, although still significant, when all data were considered (Fig. 9A) but was not significant for the Time 1 dataset (Fig. 9B). Again, these distributions were not clearly linked to larval swimming velocities, although some patterns emerged when taxonomy was considered. During frontal occurrence, barnacle cyprids and decapod mean depths were around 10m, but these ranged from 5 to more than 20m after the front has passed. Also, at Time 1 nauplii, bryozoan cyphonauts and echinoderms were at depths shallower than 10m, associated to shorewards flow. At Time 2, the same taxa remained at very similar depths, even though mean shorewards velocities were much lower than during Time 1, or even negative (Fig. 9).

A strong correspondence was found between the taxon specific Pc1 scores of the PCA for the whole dataset and the goodness of fit between larval and V vertical distributions, measured as the $R^2$ of their corresponding regressions. Thus, the main component structuring larval community was clearly correlated with the degree of coupling between larval vertical distribution and cross shore flow vertical profiles (Fig. 10A). This pattern was not associated with organismal velocity, with slow swimmers like bryozoans, gastropod and bivalve veligers presenting different values of both $R^2$ and Pc1. On the other hand, athough not significant, this degree of coupling was associated with the mean cross shore flow (Fig. 10B). Thus, the better the coupling between larval vertical positioning and shorewards flow structure, the higher the shorewards currents experienced by larvae.

Discussion

Field observations show that fronts developed off Sardinia Bay in austral spring of 2014 and they affected the distribution of coastal invertebrate larvae. Specifically, fronts were conspicuous as consecutive foam lines or oily slicks appearing parallel to the shore during spring tides as water level was close to its mínimum. Significant surface accumulations of larvae were, however, not detected at the foam lines, in agreement with low, or even negative, accumulation speeds around the structure. On the other hand, 3-4 hours later, during rising spring tides, fronts disappeared and strong shoreward flows developed at mid to bottom depths. This coincided with the presence of high larval abundances at the same depths, in marked contrast with their previous vertical distributions during frontal occurrence. Such a shift in vertical distributions is consistent with the correspondence between larval and cross shore flow mean depths for many taxa. High correlation observed between larval mean depths and larval mean shorewards currents (V) were only observed after frontal occurrence, however, so that V was important in explaining larval community assemblages, but only after frontal dissipation.

Physical forcing driving frontal development

The timing of appearance of both surface fronts and deeper strong shoreward currents point to tidally forced internal waves as the most likely drivers of the structures observed during our cruises. Alongshore foam lines are consistent with internal tidal bore fronts which move onshore as inshore subthermocline waters start to receed and sink (Pineda 1994, Pineda 1999, Weidberg et al. 2014). Fast shorewards mid to bottom currents typically occur at the following phase of wave development, as the next set of incoming waves approach inshore waters (Pineda 1991, Leichter et al. 1998). This temporal pattern reflects the dynamics of the internal baroclinic tide. Although baroclinic tides do not have to be in phase with surface barotropic tidal variations of sea level, the onset of strong subthermocline landward flows can coincide with rising tides, especially if these are marked (Leichter et al. 1998), thus being in opposite phase compared with usual estuarine circulation. We found the same association between these strong tidal currents and the rising tide when these motions are supposed to be stronger, that is, during spring tides (Fig. 2). The lack of strong winds observed during sampling is also consistent with the known wind speed range at which these structures form and persist, i.e. not more than $10 m*s^{-1}$ (Soloviev and Lukas 2014). Thermal stratification and the existence of an interface between warm surface waters and a cooler subthermocline layer are, however, required for internal wave generation and transmission (Helfrich and Pineda 2003). During our sampling, thermal vertical gradients were much lower ($0.1°C*m^{-1}$, Fig. 3) than those measured in coastal waters off Southern California when conspicuous internal waves developed ($0.9°C*m^{-1}$, Pineda et al 1999). The thermal values found during this study are in the stratification range observed off Florida and California, when internal waves were also observed (Leichter et al. 1998, Noble et al. 2009, Ryan et al. 2014), although the typical chlorophyll a maximum associated with the thermocline was marked in those studies, but not in our samplings (Fig. 3, 4 and 5). Nevertheless, Largier (1994) observed internal tides and propagation of slicks over the shelf off Cape Town when the water columnn was less stratified (around $0.04°C*m^{-1}$) and hypothesised that, under the proper topographic characteritics, the internal tide can be enhanced by the resonance between several generation sites on the shelf break, despite a low thermal stratification background. Similarly, internal tides were detected off the Western coast of Scotland across slightly stratified water columns ($0.02°C*m^{-1}$) and were thought to be intensified by coastal trapped waves (Rippeth and Inall 2002). Coastal trapped waves are known to propagate eastwards along the Agulhas Bank (Schumann and Brink 1990) and could exert similar effects on the internal tide along this coast. Thus, the development of internal waves across slightly stratified water columns may have ocurred during our sampling. In fact, this is the most plausible explanation for the slicks observed in the field and in LANDSAT images on Event 2 (Supplementary Information 1). These images revealed a banded pattern in contrast, consisting in 5-6 lines separated aproximately 100m from each other. Similar structures have already been observed in detail off Monterey Bay in California using satelite imagery similar to the one used in our study and interpreted as wind rows or Langmuir cells (Ryan et al. 2010). Langmuir cells are also characterised by parallel foam lines caused by winds acting on surface waters, but zonal winds responsible of Ekman transport were not blowing during Event 2 (Fig. 3). If Langmuir circulation was occurring, only alongshore winds could have formed the shore parallel lines observed as these usually develop in parallel to wind direction, but alongshore wind speeds were close to zero on that particular date. Furthermore, these lines were observed moving onshore, and Langmuir foam lines are known to remain at fixed locations (Soloviev and Lukas 2014). LANDSAT images also revealed a different structure further offshore consistent in an alongshore single line that was not sampled during Event 2 (Supplementary Information 1). Such a feature could correspond to a shear front due to flow separation at the tip of Cape Recife or to an upwelling shadow front (McCabe et al. 2006, Ryan et al. 2010).

Larval distributions affected by internal motions

Two modes of horizontal, landward transport of zooplankton mediated by internal waves have been described: surface accumulation at convergent slicks formed between the upwelled inshore and warmer offshore water masses (Shanks 1983, Pineda 1994, Pineda 1999, Vargas et al. 2004, Weidberg et al. 2014) and subthermocline transport by tidal currents during the next phase of wave development (Pineda 1991, Leichter et al. 1998). These two modes are not mutually exclusive, as surface and bottom transport of barnacle nauplii and crab zoeae, respectively, have been observed during opposite phases of internal wave development (Lievana MacTavish et al. 2016). Our data, however, clearly support the subthermocline transport mode and not surface accumulation at the front. Surface convergent flows were relatively weak or even negative (Table 1), thus failing to transport larvae to the front, while both strong surface convergent currents and planktonic aggregations have been found in other systems (Pineda 1999, Weidberg et al. 2014). On the other hand, much higher accumulation speeds were observed at the bottom layers during Time 2 for Events 1 and 2 (Table 1), as strong subthermocline shorewards flow occurred during the rising tide and after frontal development (Figs. 4, 5 and 6). These currents reached up to 90 cm*s$^{-1}$ (Fig 4 and 5) and may provide fast onshore advection of larvae at those depths. In agreement with these observations, both nearshore measurements and model simulations show that very strong currents can develop close to the bottom as internal waves shoal and break (Aghsaee et al. 2010, Richards et al. 2013).

Processes affecting nearshore larval distributions are often explored in a lagrangian framework, assuming that internal waves may transport offshore larvae that were previously outside the domain sampled to the coast. An horizontal eulerian perspective can however also be considered, assuming that most of the larvae sampled after the occurrence of the fronts were already there when the fronts were present and that the change in their distribution is not due to the arrival of offshore larvae, but to a shift in the vertical positioning of local larvae. In fact, our sampling design allows a comparison between larval abundances during and after the presence of fronts. Thus, for all the taxa which were located deeper in the water column at Time 2 for Events 1 and 2, larval abundances were overall not clearly higher in Time 2 compared to Time 1 (Table 2), which points to a minor role of advection from outside our domain. More likely, the vertical distribution of local larvae was altered from Time 1 to Time 2. This perspective has been considered for holoplanktonic organisms at offshore waters, but also in lakes, where vertical oscillations of the pycnocline match vertical displacements of the distribution of the zooplankters (Mc Manus et al. 2005, Rinke et al. 2007, Huber et al. 2011, van Haren 2014). Such a correspondence beween larval vertical patterns and internal motions may indicate that the organisms are passively sinking and rising with the transmission of the internal waves. The physical data indicate that there was not a clear stratified water column with a pycnoline that larvae could follow passively (Fig. 3 and 4) and, in fact, while the fronts were present on Events 1 and 2, most larval taxa were not at the surface, but evenly distributed in the vertical axis (Table 2). It was only after frontal occurrence that larvae sank and strong shorewards tidal currents developed. This points to active organismal behaviour.

Active depth regulation in response to dynamic flows

The role of vertical migration in the spatial and temporal distributions of planktonic organisms has been described in many locations. The most studied type of migration observed in many different coastal and oceanic systems across multiple taxa is diel vertical migration. Our results point, however, to a more dynamic, flexible vertical migration in response to shoreward flows. Such behaviour may not be incompatible with diel vertical migration given the highly adaptive nature of larval behaviour across the water column, but it might prevail for coastal meroplankton under a contrasting vertical structure in cross shore flow. The correpondence between larval mean depths and cross shore flow vertical profiles (Fig. 7), the strong effect of these northward flows in structuring the larval community, especially when they were fast and correlated with depth during Time 2 (Fig.8, Fig. 9), and finally the association between larval community assemblages and the larval-onshore flow coupling for each taxon (Fig. 10) may indicate that these organisms were performing highly adaptable vertical movements in search of shorewards transport. Moreover, such adaptive behaviour was successful in terms of onshore advection as those taxa exhibiting the best larval mean depth- shorewards mean depth couplings tended to experience stronger shorewards flows (Fig. 10B). From the perspective of selective pressure, behaviours enhancing the arrival of larvae at a relatively narrow and reduced adult habitat on the shore in a highly advective pelagic realm may be strongly favoured and become more important than thermal and productivity characteristics of the water masses (Crisp 1976, Sulkin 1984, Naylor 2006). For larvae, in sharp contrast with pelagic holoplankton (Jackson and Strathman 1981), to grow in productive waters with a suitable thermal range may not enhance survival if by the end of development they are not in the adult habitat on the coast. This is consistent with the fact that the distributions of mussel larvae over the Agulhas Bank were uncorrelated with temperature (Porri et al. 2014). In this context, the greater importance of cross shore currents over environmental hydrographic variables is consistent with the selective pressures to which coastal meroplankton are subjected. In addition, complex behaviours that influence retention within those water layers moving onshore might have evolved. Certainly, when larval distributions were examined at the inshore waters with respect to the structure of cross shore currents, the existence of such behaviours became clear for several taxa. Essentially, they consist of vertical migration against vertical currents under Ekman and tidal forcing (Cronin 1982, Queiroga and Blanton 2004, Genin et al. 2005, Shanks and Brink 2005, Shanks and Shearman 2009). By opposing vertical currents, larvae would remain in those layers advected to the shore in each scenario. Nevertheless, during our cruises, vertical currents did not present a clear pattern across the water column (Fig. 3, 4 and 5). Biological rhytms in phase with the tide are less likely in our coastal site, as, in contrast with tidal channels and estuaries, the synchronisation between barotropic and baroclinic tides at the open coast may not occur or may even be in opposite phase to the expected (Leichter et al. 1998). This mismatch would make the evolution of internal biological synchronisation to tides very unlikely outside of estuaries. Thus, the underlying mechanisms behind the association between larval and cross shore currents vertical profiles remain unclear and they cannot be fully inferred from this observational field study. Further experimental work is required to directly evaluate larval vertical responses to contrasting flows under different physical conditions. The inclussion of coastal sound in such studies may be of particular interest, as it could be perceived by a wide range of zooplankters and it clearly indicates onshore direction (Rogers and Cox 1988, Radford et al. 2007, Lillis et al. 2014).

Role of swimmimg ability

Across surface frontal convergences, spatial gradients in the planktonic community can be observed driven by larval swimming speed with respect to horizontal and vertical convergent flows (Franks 1992, Pineda 1999, Hetland et al. 2002, Weidberg et al. 2014). Thus, relatively fast zooplankters may accumulate as they can oppose downward currents at the front, while slower organisms may sink and recirculate mainly towards the stratified side of the structure. Our results, however, suggest that swimming speed did not influence downward migration nor the degree of MLD-MVD coupling or larval community assemblages (Fig. 6 and 7), probably because there was no persistent vertical flow opposing larval movement (Fig. 3, 4 and 5). This is in agreement with observations of larval transport by deep tidally driven shorewards flows, a process which affects different types of zooplankters, regardless of their swimming speeds (Pineda 1991, Leichter et al. 1998). On the other hand, at surface convergences, the water is forced downwards at speeds that can overcome horizontal flows (Kilcher and Nash 2010, Weidberg et al. 2014) and can pose a constraint to larval motion, leading to spatial dislocation of populations based on relative swimming speed (Franks 1992).

Conclusions

During our coastal cruises, tidally forced internal wave dynamics were observed despite low levels of thermal stratification. The role of larval onshore transport by surface fronts was minor compared to that driven by strong deep shorewards currents during the opposite phase of the baroclinic internal tide. The clear association of many different larval taxa with these flows most likely indicates an active, finely tuned migration pattern through the water column which may allow landward transport, at least for local larvae from the inner shelf to the offshore side of the surf zone. In contrast to surface convergences, larval swimming speed was not related to this process and it occurred during the phase of the internal tide when surface fronts were not present. In order to infer the role of internal tidally-forced motion in larval transport, the measurement of frequency, strength and persistence of surface manifestations of fronts, as foam lines or slicks, is of at least the same relevance as a full characterisation of these deep shorewards, post-frontal currents.

Acknowledgements

This research was funded by the African Coelacanth Ecosystem Programme (ACEP III) and it is supported by the South African Research Chairs Initiative of the Department of Science and Technology, the National Research Foundation and the South African Institute for Aquatic Biodiversity. The South African Weather Service provided valuable wind data. We thank Rachel Ndhlovu for her work in larval identification and the captain and crew of the RV uKwabelana, Ryan Palmer, Jacqui Trasierra, Shana Mian, Olwethu Duna, Carlota Fernández for assistance and data collection.

Figure legends

Figure 1. Maps of the study site. A scale of sea surface temperature is presented for the general maps. In the high resolution Google Earth map, all sampling stations are shown. The white line delimits the marine protected area of Sardinia Bay.

Figure 2. Time series of tides and zonal Ekman transport for the three events on 25-09-2014 (A), 6-10-2014 (B) and 29-10-2014 (C). Red and blue rectangles show the times at which nearshore waters were sampled during and after the presence of fronts, respectively. On the X axis, dark and yellow lines mark night-time and daytime periods, respectively.

Figure 3. Contour profiles of chlorophyll-a, temperature and currents along the 3 axes for Event 1, along two transcects (A and B), during (T1) and after frontal development (T2). Dark inverted triangles mark show the position of each sampled station.

Figure 4. Contour profiles of chlorophyll-a, temperature and currents along the 3 axis for Event 2 along two transcects (A and B) during (T1) and after frontal development (T2). Dark inverted triangles mark show the position of each station.

Figure 5. Contour profiles of chlorophyll-a, temperature and currents along the 3 axis for Event 3 along two transcects (A and B) during (T1) and after frontal development (T2). Dark inverted triangles mark show the position of each station.

Figure 6. Examples of larval distribution around nearshore frontal structures: (A) annelid segmented larvae at Event 1 after frontal development; (B) bryozoan cyphonauts at event 3 during frontal development; (C) balanid late nauplii at Event 2 during frontal development. Letters show significant groupings according to the corresponding analyses of variance, see Table 2 for levels of significance and acronyms in panel C.

Figure 7. Examples of OLS regressions between larval and meridional flow (V) mean depths for brachyuran zoeae (A, $R^2=0.69$, $P<0.0001$), bivalve veligers (B, $R^2=0.65$, $P<0.0001$) and bryozoan cyphonauts (C, $R^2=0.06$, $P=0.28$). Vertical current profiles of V (D) and distributions of balanid early nauplii (E), chthamalid early nauplii (F) and gastropod veligers (G) are shown for the offshore station A during Event 1 in Times 1 and 2 (red and blue line, respectively). The dashed line shows the limit between shorewards and seawards flows in panel D.

Figure 8. PCA analyses for larval taxa for all samples (A), for time of frontal development, Time 1 (B) and after front occurrence, Time 2 (C). Each point represents a taxon, see legend for taxonomic groups and larval velocities. Percentage of variability is shown for each PC axis along with the linear coefficient of the most important variable (T=temperature, CHLA= chlorophyll-a, U=zonal flow, V=meridional flow, Z=vertical flow).

Figure 9. Regressions between larval mean depth and mean cross shore flows for all sampling occasions (A), during frontal development, Time 1 (B) and after frontal occurrence, Time 2 (C). Each dot is a taxon, see legend for general taxonomic groups and typical organismal velocities. Significance for regression is shown (*: 0.05>p>0.01; **: 0.01>p>0.001; ***: 0.001>p)

Figure 10. Linear relationships of the taxon-specific $R^2$ between larval and meridional flow mean depths with Pc1 score (A) and mean shorewards current velocities (B). Each dot is a taxon, see legend for general taxonomic groups and typical organismal velocities.

Table 1. Accumulation speeds in cm/s from the onshore and offshore sides for each frontal event, during (Time 1) and after (Time 2) development. Bold letters indicate positive accumulation speeds were obtained. Temporal differences (T2-T1) are also shown.

| Event | Depth | Time 1 | Δvon | Δvoff | Time 2 | Δvon | Δvoff | Δvon(T2-T1) | Δvoff (T2-T1) |
|---|---|---|---|---|---|---|---|---|---|
| 1 | Surface | | **1.177** | -7.6725 | | -1.062 | **2.2855** | -2.239 | 9.958 |
| | Middle | | **1.2875** | -6.831 | | -0.47 | -0.223 | -1.7575 | 6.608 |
| | Bottom | | -18.575 | **7.117** | | -2.9545 | **35.146** | 15.6205 | 28.029 |
| 2 | Surface | | **3.937** | **0.5125** | | -1.9975 | **5.205** | -5.9345 | 4.6925 |
| | Middle | | **7.123** | **0.2575** | | -1.904 | **0.258** | -9.027 | 0.0005 |
| | Bottom | | -44.21 | -1.0065 | | **3.526** | **20.603** | 47.736 | 21.6095 |
| 3 | Surface | | **8.1025** | **1.578** | | **3.31** | -1.523 | -4.7925 | -3.101 |
| | Middle | | **0.1365** | **3.8415** | | **2.063** | **1.528** | 1.9265 | -2.3135 |
| | Bottom | | -6.0725 | **4.2325** | | -4.09 | -2.0335 | 1.9825 | -6.266 |

Table 2. Results of the analyses of variance performed on each larval taxon at each event at the time of front occurrence and after (Times 1 and 2, respectively). Significant effects are represented (*: 0.05>p>0.01; **: 0.01>p>0.001; ***: 0.001>p). Blank cells correspond to non-significant terms. Cells containing "no" show that a given taxon was not present at that event and time. Mean abundances and their standard deviations at Times 1 and 2 are shown in white for all taxa. If interactions containing any individual factor were significant, the significance of those factors on their own were not taken into account. Besides signifcant terms, significant differences are shown between factor levels (on=onshore, f=front, off=offshore; s=surface, m=mid depths, b=bottom). For the interactions, different combinations of levels are possible (i.e. off-m= offshore mid depths) and only the combinations with the highest larval abundances are shown. The term "rest" refers to the remaining combinations of levels product of the statistical interactions not showed. Significant differences found among levels contained within the term "rest" are not represented.

| Taxa | TIME 1 | Position | Depth | Position*Depth | TIME 2 | Position | Depth | Position*Depth |
|---|---|---|---|---|---|---|---|---|
| EVENT 1 | | | | | | | | |
| Balanid early nauplii | 350.41±480.91 | | | **; off-m>rest | 348.77±989.66 | | **;m>b=s | |
| Chthamalid early nauplii | 5.86±16.52 | | | | 6.97±13.99 | | *; m≥b≥s | |
| Balanid late nauplii | 4.24±6.32 | | | *;on-b=f-b=f-m=off-m>rest | 9.58±27.9 | | **;m>b=s | |
| *O. angulosa* late nauplii | 0.35±0.89 | | | | | no | no | no |
| Balanid cyprid A | 1.09±1.53 | | | | 1.04±1.56 | | *; m≥b≥s | |
| Balanid cyprid B | 1.31±1.76 | | | | 1.33±2.71 | | | *;off-b=off-m>rest |
| Brachyuran zoeae | 3.56±5.15 | | *; b=m>s | | 1.69±2.34 | **;off>f=on | ***;m=b>s | |
| Brachyuran megalopae | 0.57±1 | *;f=off>on | **;m>b=s | | | no | no | no |
| Porcelanid zoeae | 2.2±4.71 | | | *; off-m>rest | 1.92±3.98 | *;off≥f≥on | | |
| Anomuran zoeae | 5.15±8.55 | | *; m≥b≥s | | 1.64±3.29 | **;off≥f≥on | *; m≥b≥s | |
| Pinnotherid zoeae | 1.73±3.07 | | | | 0.93±1.53 | | | |
| Other annelids | 79.28±134.18 | | **; m=b>s | | 51.08±73.31 | *;off≥f≥on | **;m>b=s | |
| Spionid annelid | 0.93±1.76 | | *;m≥b≥s | | 0.2±0.34 | | | |
| Gastropod veligers | 27.34±27.2 | | | **; m=b>f-s>rest | 28.56±45.61 | **;off>f=on | ***; m>b=s | |
| Bivalve veligers | 139.65±146.12 | | ***; m=b>s | | 122.13±196.17 | | ***; m>b=s | |
| Bryozoans cyphonauts | 4.76±6.1 | **;on≥f≥off | *; m=s>b | | 3.05±4.26 | | **;m=s>b | |
| EVENT 2 | | | | | | | | |
| Balanid early nauplii | 40.01±135.76 | | | | 44.21±81.35 | *;on≥off≥f | *;m>b≥s | |
| Balanid late nauplii | 0.9±1.46 | | | *;f-m=off-m>rest | 10.12±26.1 | | | *;off-m>rest |
| Balanid cyprid A | | no | no | no | 0.86±1.39 | | **;b>m=s | |
| Balanid cyprid B | 1.82±2.55 | | | *;on-m=b>rest | 1.06±1.91 | *;on≥f≥off | | |
| Brachyuran zoeae | 3.68±7.71 | | | | 4.58±8.28 | *;on>f≥off | **;b>m=s | |
| Porcelanid zoeae | 0.48±0.94 | | | **;off-b=off-m>rest | 0.44±0.9 | | | |
| Anomuran zoeae | 1.04±1.78 | | *;b≥m≥s | | 0.84±1.29 | | *;b>m=s | |
| Pinnotherid zoeae | 0.87±2 | | | | 1.68±3.02 | *;on>f≥off | *;b>m=s | |
| Other annelids | 197.04±230.96 | | ***;b>m=s | | 147.14±213.1 | | **;b≥m≥s | |
| Gastropod veligers | 30.11±60.76 | | | *;b=on-m>rest | 26.35±54.35 | | **;b≥m≥s | |
| Bivalve veligers | 49.33±63.66 | *;on≥off≥f | **;b≥m≥s | | 73.66±43.95 | | *;b=m>s | |
| Bryozoans cyphonauts | 0.57±1.44 | | | **;on-m>rest | | no | no | no |
| EVENT 3 | | | | | | | | |
| Balanid early nauplii | 141.17±364.45 | | | | 205.82±422.21 | | | |
| Chthamalid early nauplii | 1.01±2.99 | *;off>f=on | | | 3.44±5.82 | | | |
| Balanid late nauplii | 6.67±19.4 | **;off>f=on | | | 7.74±12.88 | | *;m≥s≥b | |
| *O. angulosa* late nauplii | 0.36±1.16 | *;off>f=on | | | 1.1±2.27 | | | |
| Balanid cyprid A | 5.15±9.19 | **;off>f=on | **;b≥m≥s | | 3.71±3.4 | | *;m=b>s | |
| Balanid cyprid B | 1.2±1.49 | | | *; off-m=b>rest | 0.92±0.98 | | | |
| Brachyuran zoeae | 5.26±5.81 | | *;b≥m≥s | | 4.57±4.63 | | | |
| Porcelanid zoeae | 1.72±3.37 | | *;b>m=s | | 1.8±2.59 | | | |
| Anomuran zoeae | 8.08±12 | *;off≥f≥on | *;b≥m≥s | | 6.91±6.63 | | | *; m=b=off-s>rest |
| Pinnotherid zoeae | 0.98±1.61 | | *;b≥m≥s | | 1.8±4.64 | | | |
| Other annelids | 17.79±44.17 | | | | 33.41±44.69 | | | |
| Spionid annelid | 1.96±4.16 | | | **; off-m>off-b>rest | 2.07±2.86 | | **;m≥s≥b | |
| Gastropod veligers | 12.2±31.34 | *;off>f=on | *;b≥m≥s | | 19.7±44 | *;off≥f≥on | | |
| Bivalve veligers | 99.08±275.69 | **;off>f=on | *;b=m>s | | 88.42±171.24 | *;off≥f≥on | | |
| Bryozoans cyphonauts | 10.16±14.5 | | **;s≥m≥b | | 8.42±11.13 | *;on≥f≥off | ***;s>m=b | |
| Brittle star pluteus | 1.12±1.83 | | | | 1.12±1.83 | | | |